\documentstyle[preprint,aps]{revtex}

\begin{document}

\draft


\title{Classical and Quantum Instantons \\
in Yang-Mills Theory in the Background of de Sitter Spacetime}

\author{Hongsu Kim and Sung Ku Kim}

\address{Department of Physics\\
Ewha Women's University, Seoul 120-750, KOREA}

\date{December, 1997}

\maketitle

\begin{abstract}
Instantons and their quantisation in pure Yang-Mills theory formulated
in the background of de Sitter spacetime represented by spatially-
closed ($k=+1$) Friedmann-Robertson-Walker metric are discussed.
As for the classical treatment of the instanton physics, first, 
explicit instanton solutions are found and next, quantities like
Pontryagin index and the semiclassical approximation to the
inter-vacua tunnelling amplitude are evaluated. The Atiyah-Patodi-
Singer index theorem is checked as well by constructing explicitly
the normalizable fermion zero modes in this de Sitter spacetime
instanton background. Finally, following the kink quantisation
scheme originally proposed by Dashen, Hasslacher and Neveu, the
quantisation of our instanton is performed. Of particular interest
is the estimate of the lowest quantum correction to the inter-vacua
tunnelling amplitude arising from the quantisation of the instanton.
It turns out that the inter-vacua tunnelling amplitude gets enhanced
upon quantizing the instanton.
\end{abstract}




\centerline {\rm\bf I. Introduction}

It is well-known that topologically degenerate vacuum structure of
non-abelian gauge theories opened up our eyes to the profound and new
aspects of non-perturbative regime of the theories such as the physics
of instantons and the mechanism of quark confinement. Particularly,
the instanton physics in the pure Yang-Mills (YM) gauge theory formulated
in flat spacetime has been thoroughly studied in the literature [1] at least
semiclassically. In the present work, we discuss the classical and quantum
instanton physics in the pure YM theory formulated in the background of
de Sitter spacetime.
The formulation of scalar and spinor field theory (particularly their
quantum field theory) in the fixed background of de Sitter spacetime
has long been a center of interest and actually much work [3] has been
done associated with this topic. Therefore, it is somewhat curious
that relatively little attempt [2] has
been made toward the formulation of vector gauge theories particularly
that of YM gauge theory in the same de Sitter background spacetime.
And partly, this state of affair has been the motivation of the 
present work. There is, however, a remarkable feature that distingushes
the formulation of YM gauge theory in de Sitter spacetime from that
of scalar or spinor theory in the same de Sitter spacetime. Suppose
one starts with the Einstein-Yang-Mills theory in the presence of the
cosmological constant and treat both the gravity sector and the 
Yang-Mills matter sector on equal footing. As long as we restrict our
interest to instanton solutions in this system, we need to look for
solutions to (anti)self-dual equation for YM field strength. Then
what happens is that for the (anti)self-dual field strength, the
YM energy-momentum tensor vanishes identically [2] in the Euclidean
signature. This indicates that the YM (matter) field does not disturb
the spacetime geometry while the geometry still does have effect on
the YM field. As a result, the geometry, which is left intact by the
YM field, effectively serves as a ``background'' spacetime which can
be chosen somewhat at our will and here in this work, we take it to
be the de Sitter spacetime (since we included the cosmological constant).
And in this work, the metric
for this background de Sitter spacetime is chosen to be that of the
spatially-closed FRW having the Lorentzian topology of $R\times S^3$ 
(with $S^3$ being the topology of the spatial section) and the 
Euclidean topology of $S^4$. Thus it has SO(4)-symmetry and hence 
the dynamical YM field put on it should have the same SO(4)-symmetry
as well. Then noticing that the SU(2) group manifold is also $S^3$ just
like it is the case for the geometry of the spatial section of the
manifold, one may choose a ``common'' basis for both the group manifold
and the spatial section of the spacetime manifold. And this indicates
that there will be ``mixing'' between the group index in the YM field
and the frame index. Namely we can employ an analogue of the `tHooft-
Polyakov's ``hedgehog'' ans\H atz for the monopole solutions [4] in 
Yang-Mills-Higgs theory. This high
degree of ``built-in'' symmetry, then reduces the system effectively to
a one-dimensional system of a self-interacting scalar field (namely, a
kind of scalar $\phi^4$-theory) with potential of the structure of that of 
``double-well'' whose vacuum has two-fold degeneracy. Of course from
this point on, one may proceed to carry out the quantisation of the
one-dimensional scalar $\phi^4$-type reduced system as a mean to
formulate the quantum YM gauge theory in de Sitter background 
spacetime. However, since associated with this degeneracy in vacuum of
the theory, of central interest is the physics of instanton, we,
instead, explore the instanton physics of this system
in the present work. As a matter of fact, there had been some works [2]
on concrete study of YM instanton solutions in curved spacetime.
Eguchi and Freund [2] considered the YM instanton in conformally-flat
general spacetimes and Charap and Duff [2] discussed it in
(maximally-extended) Schwarzschild spacetime. The detailed account of
relationships between our present work and those of these authors will
be given at the end of sect.II and sect.III respectively. 
In classical
terms, instanton is a gauge field configuration which interpolates between
two degenerate but distinct vacua. Or more rigorously, it is a classical 
solution to the Euclidean equation of motion that makes dominant 
contribution to the inter-vacua tunnelling amplitude. As a classical
treatment of this instanton physics, first, explicit instanton solutions
will be found. Next, quantities like Pontryagin index representing the
instanton number and the semiclassical
approximation to the vacuum-to-vacuum tunnelling amplitude will be
evaluated. And lastly, the Atiyah-Patodi-Singer index theorem will be confirmed
by constructing explicitly the normalizable fermion zero modes in this
instanton background.  Then follows the quantum treatment of the instanton
physics. As will be shown in the text later on, since the Euclidean time
is just another ``spacelike'' coordinate, the Euclidean action of the
reduced one-dimensional system may be viewed as the potential energy or
the Hamiltonian of a system of ``static'', self-interacting scalar field.
As a consequence, one can directly apply the standard, conventional soliton
(particularly kink) quantisation formalism developed in original papers [5,6]
to the quantisation of our instanton. Among various quantisation techniques,
we shall employ, in the present work, that of Dashen, Hasslacher and Neveu [5].
As is well-known, in the context of this soliton quantisation scheme, the
leading quantum correction corresponds to the contribution of a set of
approximate harmonic oscillator states. Thus energy levels of quantized
instanton wll be given. Finally, as a result of central importance in this
work, we shall provide the lowest order quantum correction to the
Euclidean instanton action and hence to the vacuum-to-vacuum tunnelling
amplitude arising from the quantisation of the instanton. 
To summarize the result, the Euclidean action of the quantized instanton
is lower than that of the classical instanton. And this suggests that
in the context of quantized instanton, the inter-vacua tunnelling amplitude
gets enhanced compared to what happens in the context of classical instanton. 
This paper is organized as follows : In sect.II, general formalism for the
pure YM theory in de Sitter background spacetime is provided. In sect.III,
we give a classical treatment of the instanton physics in this system.
Sect.IV is particularly prepared for the confirmation of Atiyah-Patodi-
Singer index theorem in the context of our system. Sect.V is devoted
to the formal quantisation of our instanton and finally in sect.VI, we
summarize the results of our study.

\centerline {\rm\bf II. General Formalism}

As mentioned in the introduction, we would like to discuss the physics
of classical instanton solution in pure YM theory formulated in de Sitter
background spacetime represented by the spatially-closed ($k = +1$)
FRW-metric. Thus we begin with the action governing our system, namely
the Einstein-Yang-Mills theory in the presence of the (positive) cosmological
constant

\begin{eqnarray}
 S_{EYM} &=& \int d^{4}x\sqrt{g} [{1\over 16\pi G}R - \Lambda
 -{1\over 4g^2_{c}}F^{a}_{\mu\nu}F^{a\mu\nu}], \\
 I_{EYM} &=& \int d^{4}x\sqrt{g} [\Lambda - {1\over 16\pi G}R 
 +{1\over 4g^2_{c}}F^{a}_{\mu\nu}F^{a\mu\nu}] \nonumber
\end{eqnarray}
in Lorentzian and Euclidean signature respectively. The classical field
equations which result from extremizing the EYM theory action above
is given by
\begin{eqnarray}
&&R_{\mu\nu} - {1\over 2}g_{\mu\nu}R + 8\pi G\Lambda g_{\mu\nu} = 8\pi G T_{\mu\nu},
\nonumber \\
&&T_{\mu\nu} = {1\over g^2_{c}}[F^{a}_{\mu\alpha}F^{a\alpha}_{\nu} - {1\over 4}
g_{\mu\nu}(F^{a}_{\alpha\beta}F^{a\alpha\beta})], \nonumber \\
&&D_{\mu}[\sqrt{g}F^{a\mu\nu}] = 0
\end{eqnarray}
where we employed the convention 
$A_{\mu} = A^{a}_{\mu}(-iT^{a})$ and $F_{\mu\nu} = F^{a}_{\mu\nu}
(-iT^{a})$ (with $T^{a} = \tau^{a}/2$, $\tau^{a}$ being Pauli spin
matrices obeying the SU(2) Lie algebra $[T^{a}, T^{b}] = 
i\epsilon^{abc}T^{c}$ and the normalization
$Tr(T^{a}T^{b}) = \delta^{ab}/2$) 
in which the YM field strength and gauge-covariant 
derivative are given respectively by 
$F^{a}_{\mu\nu} = \partial_{\mu}A^{a}_{\nu} - \partial_{\nu}A^{a}_{\mu}
+ \epsilon^{abc}A^{b}_{\mu}A^{c}_{\nu}$,
$D^{ac}_{\mu} = \partial_{\mu}\delta^{ac} + \epsilon^{abc}A^{b}_{\mu}$.
$a,b,c = 1,2,3$ denote SU(2) group indices and $g_{c}$ is the YM
gauge coupling constant. \\
As mentioned in the introduction, we are particularly interested in the
solution to (anti)self-dual equation in Euclidean signature to find the
de Sitter spacetime version of the YM instanton. Then the Euclidean YM
field energy-momentum tensor vanishes identically, $T_{\mu\nu} = 0$ and
the Einstein field equation reduces to that of de Sitter spacetime,
$R_{\mu\nu}-g_{\mu\nu}R/2+8\pi G\Lambda g_{\mu\nu}=0$. In order to
represent this de Sitter spacetime, now we employ the spatially-closed
($k=+1$) FRW-metric given by
\begin{eqnarray}
 ds^{2} &=& g_{\mu\nu}dx^{\mu}dx^{\nu} =
            \eta_{AB} e^{A}{\otimes}e^{B} \\
 &=& [-N^{2}(t)dt^{2} + a^{2}(t)\sigma^{a}{\otimes}\sigma^{a}] \nonumber \\
 &=& [N^{2}(\tau)d{\tau}^{2} +
 a^{2}(\tau)\sigma^{a}{\otimes}\sigma^{a}] \nonumber
\end{eqnarray}
in Lorentzian and Euclidean signature respectively and
where $N(t)$ and $a(t)$ are lapse function and scale factor
respectively. First in the gravity sector (which is left intact by the 
YM field), there is a gauge arbitrariness
which amounts to the invariance of the curvature
under the 4-dim. diffeomorphisms, i.e., general coordinate transformations.
And this 4-dim. diffeomorphism consists of the time-reparametrization
corresponding to possible different choices for the lapse function $N(t)$
and the 3-dim. diffeomorphism corresponding to the freedom in choosing
coordinates for the left-invariant basis 1-forms $\{\sigma^{a}\}$ 
representing the metric on the spacelike hypersurface $S^3$. Here in this
work, our choice for the gauge-fixing will be determined as follows :
since this background spacetime metric has the Lorentzian topology of 
$R\times S^3$ and the Euclidean topology of $S^4$,
it has SO(4)-symmetry. Thus in order to take advantage of this high
degree of symmetry, we shall employ the Euler angle coordinates
($\theta, \phi, \psi$) parametrizing the spatial section of the manifold,
$S^3$. This is the gauge fixing associated with the 3-dim. diffeomorphism.
Next, concerning the time-reparametrization freedom, we shall mainly
employ two alternative gauges $N(t) = 1$ and $N(t) = a(t)$, i.e., the
so-called ``conformal-time'' gauge. In the first gauge $N(t) = 1$, the
scale factor satisfying the Einstein equation for de Sitter spacetime
is $a(t) = {1\over \kappa}\cosh (\kappa t)$ 
($a(\tau) = {1\over \kappa}\cos (\kappa \tau)$ in Euclidean time
$\tau = it$) where $\kappa = \sqrt{8\pi G\Lambda /3}$
while in the second gauge  $N(t) = a(t)$, the corresponding
scale factor is given by $a(t) = {1/\kappa \cos t}$
($a(\tau) = {1/\kappa \cosh \tau}$ in Euclidean signature).
Next, for reasons that will become clear later on, throughout this work,
we shall mainly work with ``non-coordinate'' basis with indices
$A,B = 0,a$ ($a = 1,2,3$) rather than with coordinate basis with indices
$\mu, \nu = t, \theta, \phi, \psi$ where ($\theta, \phi, \psi$) are again
the Euler angles. The non-coordinate basis 1-forms can be read off from
the metric given in eq.(3) as 
\begin{eqnarray}
e^{A} = \{e^{0} = Ndt, ~~e^{a} = a\sigma^{a}\}
\end{eqnarray}
where $\{\sigma^{a}\}$ $(a = 1, 2, 3)$ form a basis on the 3-sphere $S^3$,
as mentioned, satisfying the SU(2) ``Maurer-Cartan'' structure
equation
\begin{eqnarray}
 d\sigma^{a} + \epsilon^{abc}\sigma^{b}{\wedge}\sigma^{c} = 0.
\end{eqnarray}
In our gauge-fixing, $\sigma^{a}$'s are represented in terms of 3-Euler 
angles $0\le\theta\le\pi$,
$0\le\phi\le2\pi$ and  $0\le\psi\le4\pi$, parametrizing $S^{3}$ as

\begin{eqnarray}
 \sigma^{1} &=& -{1\over 2}(sin{\psi}d{\theta} - cos{\psi}sin{\theta}d{\phi}), 
\nonumber\\
 \sigma^{2} &=& {1\over 2}(cos{\psi}d{\theta} + sin{\psi}sin{\theta}d{\phi}),\\
 \sigma^{3} &=& -{1\over 2}(d{\psi} + cos{\theta}d{\phi}).\nonumber
\end{eqnarray}
For later use, we also write down the associated vierbein and its inverse
using the definition, $e^{A} = e^{A}_{\mu}dx^{\mu}$ ,
$e^{A}_{\mu}e^{\mu}_{B} = \delta^{A}_{B}$ and
$e^{\mu}_{A}e^{A}_{\nu} = \delta^{\mu}_{\nu}$ where
$x^{\mu} = (\tau,\theta,\phi,\psi)$
\begin{eqnarray}
 e^{A}_{\mu} = \left(\matrix
               { N & 0          & 0                     & 0 \cr
                 0 & -{a\over 2}sin{\psi} & {a\over 2}cos{\psi}sin{\theta} & 0 \cr
                 0 & {a\over 2}cos{\psi} & {a\over 2}sin{\psi}sin{\theta} & 0 \cr
                 0 & 0          & -{a\over 2}cos{\theta}          & -{a\over 2} }
               \right)\quad,\quad
 e_{A}^{\mu} = \left(\matrix
      { {1\over N} & 0          & 0                     & 0 \cr
                 0 & -{2\over{a}}sin{\psi}
                   & {2\over{a}}cos{\psi}
                   & 0 \cr
                 0 & {2\over a}{cos{\psi}\over{sin{\theta}}}
                   & {2\over a}{sin{\psi}\over{sin{\theta}}}
                   & 0 \cr
                 0 & -{2\over a}{cos{\psi}cos{\theta}\over{sin{\theta}}}
                   & -{2\over a}{sin{\psi}cos{\theta}\over{sin{\theta}}}
                   & -{2\over{a}} }
               \right).
\end{eqnarray}
Thus far we have discussed the choice of ans\H atz for the metric (i.e., $k = +1$
FRW-metric which is SO(4)-symmetric) and the gauge-fixing for gravity sector.
Next, we turn to the choice of ans\H atz for the YM gauge potential and the 
SU(2) gauge-fixing. And here, our general guideline is that since the background
de Sitter spacetime metric is chosen to possess SO(4)-symmetry, the dynamical
YM field put on it should have the SO(4)-symmetry as well. Then next, note that
the SU(2) group manifold is also $S^3$ just as it is the case for the geometry
of the spatial section of the spacetime manifold. Thus one may choose the
left-invariant 1-form $\{\sigma^{a}\}$ as the ``common'' basis for both the 
group manifold and the spatial section of the spacetime manifold. And this 
indicates that there is now ``mixing'' between the group index in the YM field
and the non-coordinate frame basis index since we, as mentioned earlier, choose
to work with non-coordinate basis. Now an appropriate choice of YM gauge
potential ans\H atz incorporating all of these conditions is [7]
\begin{eqnarray}
A^{a} = A^{a}_{\mu}dx^{\mu} = [1 + H(t)]\sigma^{a}.
\end{eqnarray}
Of course in taking this YM gauge potential ans\H atz, we implicitly chose
the ``temporal gauge-fixing'' $A_{t} = 0$ (or $A_{0} = 0$ in non-coordinate
basis). This gauge choice is indeed natural since the background spacetime 
metric is homogeneous and isotropic thus depends only on time coordinates,
there is no gauge freedom associated with the space-dependent gauge
transformation. By now, it should be clear that it is more appropriate to
work with non-coordinate basis. And in the formulation employing the use
of non-coordinate basis, various equations involved should be put in 
differential forms. To be more specific, the definition for the YM field
strength takes the form, 
$F^{a} = dA^{a} + {1\over 2}\epsilon^{abc} A^{b}\wedge A^{c}$ which,
using the gauge potential ans\H atz above, is computed to be
\begin{eqnarray}
F^{a}_{0b} &=& {\dot{H}\over Na}\delta^{a}_{b}, \\
F^{a}_{bc} &=& {(H^2 - 1)\over a^2}\epsilon^{abc} \nonumber
\end{eqnarray}
where ``dot'' denotes the derivative with respect to Lorentzian time $t$.
Next, the classical YM field equation and the Bianchi identity are
\begin{eqnarray}
D\tilde{F} &=& d\tilde{F} + A\wedge \tilde{F} - \tilde{F}\wedge A, \\
DF &=& dF + A\wedge F - F\wedge A \nonumber
\end{eqnarray}
respectively with ``tilde'' denoting the Hodge dual. As far as the classical 
treatment of the system is concerned, one is mainly interested in solving
the classical YM field equation. Thus particularly associated with the
instanton physics that we shall discuss later on, it seems worth noticing
that in the pure YM theory, the solutions to (anti) self-dual equation
\begin{eqnarray}
F^{a} = \pm \tilde{F}^{a},
\end{eqnarray}
which is just 1st order differential equation, are automatically solutions of
the classical YM field equation (owing to the Bianchi identity) as well as the
minima of the Euclidean YM theory action. Another point to mention is that
the SO(4)-symmetric ans\H atz for the YM gauge potential chosen above does
obey the Bianchi identity as it should. For later use, we write down the
(anti) self-dual equation above in terms of the SO(4)-symmetric ans\H atz 
for the metric and the YM gauge potential
\begin{eqnarray}
{H'\over Na} = \pm {(H^2 - 1)\over a^2}
\end{eqnarray}
where now ``prime'' means the derivative with respect to the Euclidean time 
$\tau$. In order to have a qualitative insight into our system prior to all
quantitative calculations, we first rewrite the action of our system
given earlier in terms of the SO(4)-symmetric ans\H atz for the background
spacetime metric and the YM gauge field. Thus using
\begin{eqnarray}
(F^{a}_{\mu\nu})^2 = (\eta^{AC}\eta^{BD}F^{a}_{AB}F^{a}_{CD}) =
6 [-({\dot{H}\over Na})^2 + ({H^2-1\over a^2})^2]
\end{eqnarray}
it follows
\begin{eqnarray}
S_{YM} &=& {r^2_{0}\over 2}\int dt [({a\over N})\dot{H}^2 - ({N\over a})
(H^2-1)^2] \\
&=& {r^2_{0}\over 2}\int d\tilde{t} [({dH\over d\tilde{t}})^2 - (H^2-1)^2]
\nonumber
\end{eqnarray}
where we introduced $d\tilde{t} = (N/a)dt$ and defined 
$r^2_{0} = {6\pi^2/g^2_{c}}$. 
Note that this pure YM system put in the background of de Sitter space
represented by the spatially-closed ($k = +1$) FRW metric has been
reduced to a one-dimensional system of a particle of unit mass with
the potential given by
\begin{eqnarray}
\tilde{V}(H) = {1\over 2}\tilde{U}(H) = {r^2_{0}\over 2}(H^2 - 1)^2
\end{eqnarray}
which has the ``double-well'' structure. 
Here and henceforth we introduce the notations for the ``potential'',
$U(H) \equiv (H^2-1)^2 = 2V(H)$ and $\tilde{U}(H)=r^2_{0}U(H)=2\tilde{V}(H)$.
Since the minimum of the potential,
i.e., the vacuum has two-fold degeneracy at $H = \pm 1$, readily we
anticipate possible quantum tunnelling phenomenon between the two degenerate
vacua. Thus in order to study this vacuum-to-vacuum tunnelling, we reformulate
this system in Euclidean time obtained by the Wick rotation
$\tilde{\tau} = i\tilde{t} = i\int dt({N\over a}) = \int d\tau ({N\over a}).$
Then the Euclidean action is given by
\begin{eqnarray}
I_{YM} = -i S_{YM} = {r^2_{0}\over 2}\int d\tilde{\tau}[H'^2 + U(H)]
\end{eqnarray}
where again the prime denotes the derivative with respect to the Euclidean 
time $\tilde{\tau}$ while the overdot we used earlier denotes that with respect to
the Lorentzian time $t$. Upon extremizing this action with respect to the
field $H$ representing the YM gauge field we now get the Euclidean equation
of motion
\begin{eqnarray}
H'' = {1\over 2}{\partial U\over \partial H}.
\end{eqnarray}
It is well-known that even without explicitly solving this equation of 
motion, one can easily determine the qualitative features of the solution
which will turn out to be the instanton solution. Thus to do so, we
consider the ``first integral'' of the Euclidean equation of motion given 
above
\begin{eqnarray}
{1\over 2}H'^2 - {1\over 2}U(H) = E
\end{eqnarray}
where $E$ is the integration constant. This first integral equation describes
a system of unit mass particle with total energy $E$ moving in the ``inverted''
potential $-V(H) = -{1\over 2}U(H)$.
Obviously, the motion of particle with zero total energy, $E = 0$, which is
of our interest, will be that the particle starts (say, at $\tilde{\tau} =
-\infty$) on top of one hill and moves to the top of the other (at
$\tilde{\tau} = +\infty$). Since this behavior of the particle (with position
$H$) in this mechanical problem corresponds to the behavior of the solution
$H(\tau)$ of the Euclidean equation of motion, we can expect that there
will be a solution of instanton type. For example, the solution of the
equation of motion satisfying the boundary condition
$\lim_{\tau \rightarrow \mp \infty}H(\tau) = \pm 1$ and
$\lim_{\tau \rightarrow \mp \infty}H(\tau) = \mp 1$ will be instanton and
anti-instanton respectively. Moreover, in pure YM gauge theories like the
one we consider here, the (anti)self-dual equation $F^{a} = \mp \tilde{F}^{a}$
always imply the Euler-Lagrange's equation of motion owing to the Bianchi
identity. Thus we only need to solve this (anti)self-dual equation to obtain
the solutions. And as we shall see shortly, (anti)instanton solutions emerge
as explicit solutions to (anti)self-dual equation depending on the choice of
gauge for the lapse function $N(\tau)$ associated with the time reparametrization
invariance of the theory of background gravity. Note also that (anti)self-dual
equation, $(dH/d\tau) = \mp (N/a)(H^2-1)$ exactly coincides with the first
integral of the equation of motion with $E=0$,
$(dH/d\tilde{\tau}) = \mp [U(H)]^{1/2}$. In this Euclidean formulation, 
Euclidean time $\tau$ is just another spacelike coordinate and hence the
instanton can be thought of as a soliton configuration (actually this is why
'tHooft dubbed the name ``instantons'' for the Euclidean solitons). Thus for
later use, here we provide the expression for the energy of the instanton as
the ``soliton energy''. Since the Euclidean action represents the energy of
the system, the soliton energy is given by
\begin{eqnarray}
\varepsilon_{soliton} &=& I_{YM}[instanton] =
{r^2_{0}\over 2}\int^{\infty}_{-\infty}d\tilde{\tau}[H'^2 + U(H)] \nonumber \\
&=& r^2_{0}\int^{\infty}_{-\infty}d\tau ({N\over a})(H^2-1)^2
= -r^2_{0}\int^{1}_{-1}dH (H^2-1) \\
&=& {4\over 3}r^2_{0} = {8\pi^2 \over g^2_{c}} \nonumber
\end{eqnarray}
where we used the first integral in eq.(18) of the Euclidean equation of
motion with $E = 0$ and proper boundary conditions for instanton solutions.
Note that this expression for the energy of the instanton displays a generic
feature commonly shared by all soliton solutions [1], namely it is inversely
proportional to the coupling constant of the theory, $g_{c}$, and hence 
the instanton is a non-perturbative object. \\
It is well-known that the vacuum in the pure YM theory in flat spacetime
is infinitely degenerate and the degenerate vacua are classified by
the topological structure (namely they fall into different homotopy
classes). Thus one might be curious about the nature of changed vacuum 
structure in our case, i.e., now we have just two-fold degeneracy in
the vacuum at $H = \pm 1$. It turns out that the two degenerate vacua
$H = \pm 1$ in this pure YM theory formulated in the background of
de Sitter spacetime are {\it not} associated with the non-trivial
topology structure. Thus it seems worth comparing between the vacuum
structure of YM theory in flat spacetime and that in the background of
de Sitter spacetime. Firstly in the pure YM theory in flat spacetime,
the vacuum corresponds to $F_{\mu\nu} = 0$, which, in terms of the
gauge potential, is described by the ``pure gauge''
\begin{eqnarray}
A_{\mu} = -{i\over g_{c}}[\partial_{\mu}g(x)]g^{-1}(x)
\end{eqnarray}
where $g(x)\in SU(2)$ denotes the SU(2) group-valued function.
Now in Euclidean signature, the spacetime has the geometry and
topology of $R^{4}$ and hence obviously its boundary at which the 
vacuum ($F_{\mu\nu} = 0$) occurs is $S^{3}$. Thus for the vacuum,
the relevant base manifold for the SU(2) group-valued function
$g(x)$ in the expression for the pure gauge above is $S^{3}$. Then
the mapping $g(x)$ from the base manifold $S^{3}$ to the group
manifold $SU(2) \sim S^{3}$ forms a non-trivial homotopy group
$\Pi_{3}(SU(2)) = \Pi_{3}(S^3) = Z$. As a result, the YM theory 
vacuum in flat spacetime is infinitely degenerate and it consists 
of homotopically-inequivalent $n$-vacua with $n$ denoting the
``winding number''. \\
Secondly in the pure YM theory in the background of de Sitter spacetime
which is the case at hand, however, the vacuum of the theory exhibits
rather different nature. Namely, unlike the flat Euclidean spacetime,
the background de Sitter spacetime represented by the spatially-closed
($k = +1$) FRW-metric has the topology of $S^4$. 
Therefore, the $k = +1$ FRW-metric itself and the YM gauge field
defined on it are both taken to possess SO(4)-symmetry. And as we
have seen, the two degenerate vacua occur for $H = \pm 1$ at
$\tau = -\infty$ and for $H = \mp 1$ at $\tau = +\infty$.
Namely for the vacuum, the relevant base manifold for the SU(2)
group-valued function $g(x)$ now turns out to be, say, $S^{0}$
(0-sphere) consisting of two points $\{\tau = -\infty, \tau =+\infty\}$.
Thus the mapping $g(x)$ from the base manifold $S^{0}$ to the group
manifold $SU(2)\sim S^3$ forms trivial homotopy group
$\Pi_{0}(S^3) = 0$. This suggests that the two degenerate vacua
$H = \pm 1$ are not associated with the non-trivial homotopy structure.
Rather, these two vacua can be thought of as an analogue of again
two degenerate vacua existing in Wu-Yang magnetic monopole [8] in spherically-
symmetric (i.e., SO(3)-symmetric) flat spacetime in which the monopole
ans\H atz for the gauge potential and its field strength are given in
spherical-polar coordinates as
\begin{eqnarray}
A^{a}_{t} &=& A^{a}_{r} = 0, \nonumber \\
A^{a}_{\theta} &=& -{1\over g_{c}}[1-u(r)]\hat{\phi}^{a},
~~A^{a}_{\phi} = {1\over g_{c}}[1-u(r)]\sin \theta \hat{\theta}^{a}
\nonumber
\end{eqnarray}
and
\begin{eqnarray}
F^{a}_{r\theta} &=& {u'(r)\over g_{c}}\hat{\phi}^{a},
~~F^{a}_{r\phi} = -{u'(r)\over g_{c}}\sin \theta \hat{\theta}^{a}, \nonumber \\
F^{a}_{\theta \phi} &=& {[u^2(r)-1]\over g^2_{c}}\sin \theta \hat{r}^{a}.
\nonumber
\end{eqnarray}
Obviously, the vacuum here, $F^{a} = 0$, amounts to two values $u(r)=\pm 1$.
Again the vacuum of this system has two-fold degeneracy which is not of
topological origin. Therefore, we can conclude that the two degenerate vacua
$H(\tau)=\pm 1$ in our theory discussed above are of exactly the same kind.
In the introduction we mentioned that we would give an account of the relationship
between the instanton solution in our present work and that in the work of
Eguchi and Freund [2]. We will do it now. Eguchi and Freund [2] considered a
Weyl-invariant theory with the action
\begin{eqnarray}
S = \int d^4x \sqrt{g}[-{1\over 2}g^{\mu\nu}\partial_{\mu}\phi \partial_{\nu}\phi
- {1\over 12}R\phi^2 - \lambda \phi^4 - {1\over 4}F^{a}_{\mu\nu}F^{a\mu\nu}]
\nonumber
\end{eqnarray}
which is indeed invariant under the Weyl rescaling
\begin{eqnarray}
g_{\mu\nu} &=& \Omega^2(x)\tilde{g}_{\mu\nu}, ~~~x^{\mu} = \tilde{x}^{\mu}, 
\nonumber \\
\phi &=& \Omega^{-1}(x)\tilde{\phi}, ~~~F^{a}_{\mu\nu} = \tilde{F}^{a}_{\mu\nu}
\nonumber 
\end{eqnarray}
where $\phi$ is a scalar field.
Then the classical field equations which result by extremizing this action with
respect to $g_{\mu\nu}$, $\phi$ and $A^{a}_{\mu}$ admit a special (Euclidean) 
solution 
\begin{eqnarray}
g_{\mu\nu} &=& \delta_{\mu\nu}, ~~~\phi = ({2b^2\over \lambda})^{1/2}
{1\over (x^2+b^2)}, \nonumber \\
A^{a}_{\mu} &=& (A^{a}_{\mu})_{BPST} \nonumber
\end{eqnarray}
where the subscript ``BPST'' indicates the flat space instanton solution.
Now one can take advantage of the Weyl-invariance property of the theory.
Namely, since the action has the Weyl-rescaling invariance given above, 
starting from this special solution, one can generates new solutions for
every conformally-flat spacetime. For instance, de Sitter spacetime is
conformally-flat and hence one might wish to construct the instanton 
solution in de Sitter spacetime. In order to make a transit to the de 
Sitter spacetime, one needs to force the scalar field $\phi$ in this 
theory to take the constant value $(3/4\pi G)^{1/2}$. This can be achieved
by taking the conformal factor to be $\Omega(x) = (8\pi Gb^2/3\lambda)^{1/2}
1/(x^2+b^2)$. Then the action above of this theory reduces to that of
Einstein-Yang-Mills theory in the presence of the cosmological constant
$\Lambda = (9\lambda/2\pi G)$. Now if we confine our interest only to
the instanton solution in the YM sector, again the YM energy-momentum tensor
vanishes identically and hence at the classical level this reduced system
does represent a proper system to discuss the YM instanton solution in
de Sitter spacetime and it (i.e., the instanton solution) turns out to 
remain the same as the flat space instanton solution. Now the questions
is ; how do we compare this (unaffected) instanton solution in de Sitter
spacetime with ours? As a matter of fact, the Eguchi-Freund instanton
solution above which was constructed via a conformal transformation like
this is not quite the instanton solution in de Sitter spacetime in the
rigorous sense. To see this, recall that the de Sitter spacetime is the
space of (positive) constant curvature and hence is conformally-flat.
And these characteristics imply that the de Sitter spacetime has the
topology of $S^4$ in Euclidean signature. Now according to the method
of Eguchi and Freund, one starts from the ``seed'' solution and perform
a conformal transformation on it to obtain the solution in de Sitter
spacetime. Since the metric solution in the seed solution is the flat 
metric with the topology of $R^4$, the conformal tranformation can
{\it at most} turn it into the metric with the topology of $S^4-\{p\}$
with $\{p\}$ denoting the ``north pole'', not the complete $S^4$.
Thus the new solution is at most the instanton solution in the, say,
``almost de Sitter'' spacetime with the topology of  $S^4-\{p\}$ but
not of complete $S^4$. This missing point $\{p\}$, then, indicates 
that the conformally-transformed metric $g_{\mu\nu}=\Omega^2(x)\delta_{\mu\nu}$
represents a non-compact spacetime with boundary corresponding to
``points at infinity'' of $R^4$ (this last point becomes manifest
if one uses the stereographic projection). Namely, even after
the conformal transformation, the background spacetime still has
the boundary of topology of $S^3$ and thus the associated YM instanton
solution maintains the non-trivial homotopy structure of its flat
space counterpart. Our solution, on the other hand, is the YM instanton
solution in {\it genuine} de Sitter spacetime with topology of complete
$S^4$. As a result, as was pointed out earlier, the instanton solution in
our system completely lacks the non-trivial homotopy structure. To 
conclude, the solution of Eguchi and Freund and that of ours are ``two
different'' solutions which can not be related by any gauge transformation.

\centerline {\rm\bf III. Classical Instanton Solutions}

As is well-known, the classical solution to the (anti)self-dual equation
$F_{\mu\nu} = \mp \tilde{F}_{\mu\nu}$ minimizes the Euclidean YM theory
action [1],
\begin{eqnarray}
I_{YM} = \int d^4x \sqrt{g} {1\over 2g^2_{c}}Tr(F_{\mu\nu}F_{\mu\nu}) \geq
\pm \int d^4x \sqrt{g} {1\over 2g^2_{c}}Tr(F_{\mu\nu}\tilde{F}_{\mu\nu})
\nonumber
\end{eqnarray}
and thus makes dominant contribution to the vacuum-to-vacuum tunnelling
amplitude. Thus here we attempt to solve (anti)self-dual equation to find
instanton and anti-instanton solutions. In terms of SO(4)-symmetric ans\H atz
for the background metric and the YM gauge field on it, the (anti)self-dual
equation takes the form given in eq.(12) which, as mentioned, coincides with
the first integral of the Euclidean equation of motion with $E = 0$.
Now we solve this (anti)self-dual equation with different gauge choices for
the lapse function $N(\tau)$ associated with the time-reparametrization
invariance of the background gravity. 

(1) In the ``conformal-time'' gauge, $N(\tau) = a(\tau)$ : 

The (anti)self-dual equation in eq.(12) becomes, in this gauge
\begin{eqnarray}
{dH\over d\tau} = \mp (H^2 - 1)
\end{eqnarray}
which, upon integration, yields
\begin{eqnarray}
H(\tau) = - \tanh \tau
\end{eqnarray}
as a solution of the self-dual equation and
\begin{eqnarray}
H(\tau) = \tanh \tau
\end{eqnarray}
as a solution of the anti-self-dual equation. \\
Thus we have the instanton solutions
\begin{eqnarray}
A^{a} = A^{a}_{\mu}dx^{\mu} = [1 \mp \tanh \tau]\sigma^{a}
\end{eqnarray}
(where $\mp $ indicates instanton and anti-instanton respectively) in the
background de Sitter spacetime with metric
\begin{eqnarray}
ds^2 = {1\over \kappa^2}{1\over \cosh^2 \tau}[d\tau^2 + 
\sigma^{a}\otimes \sigma^{a}].
\end{eqnarray}
(2) In the gauge, $N(\tau) = 1$ : 

The (anti)self-dual equation in eq.(12) becomes, in this gauge
\begin{eqnarray}
{dH\over d\tau} = \mp {1\over a(\tau)}(H^2 - 1)
\end{eqnarray}
(with $a(\tau)={1\over \kappa}\cos (\kappa \tau)$)
which, upon integration, yields
\begin{eqnarray}
H(\tau) = - \sin(\kappa \tau)
\end{eqnarray}
as a solution of the self-dual equation and
\begin{eqnarray}
H(\tau) = \sin(\kappa \tau)
\end{eqnarray}
as a solution of the anti-self-dual equation. \\
Note here that the cosmological constant $\Lambda $ that determines
the background spacetime as the classical de Sitter spacetime should
be $\Lambda << M^2_{p}$ (where $M_{p}=G^{-1/2}$ denotes the Planck mass which
sets the lower bound for the scale for quantum gravity) since the
``background'' spacetime is supposed to be a classical gravity with
fixed geometry (and topology). This, in turn, implies that the period
of the (anti)instanton solution above is infinite, i.e.,
$(period) = 2\pi/\kappa = 2\pi/\sqrt{8\pi \Lambda /3M^2_{p}} 
\rightarrow \infty$ and hence the shape of the (anti)instanton
solution in this gauge $N(\tau) = 1$ is essentially the same as that
of the (anti)instanton solution in the previous gauge $N(\tau )=a(\tau )$.
Thus we have the instanton solutions
\begin{eqnarray}
A^{a} = A^{a}_{\mu}dx^{\mu} = [1 \mp \sin(\kappa \tau)]\sigma^{a}
\end{eqnarray}
(where $\mp $ indicates instanton and anti-instanton respectively) in the
background de Sitter spacetime with metric
\begin{eqnarray}
ds^2 = [d\tau^2 +
{1\over \kappa^2}\cos^2 (\kappa \tau) \sigma^{a}\otimes \sigma^{a}].
\end{eqnarray}
Finally, note that regardless of the gauge choice for the lapse function,
the (anti)instanton solutions are all $-1 \leq H(\tau)\leq +1$ and hence
interpolate the two degenerate vacua $H = \pm 1$ as they should. \\
Upon constructing the classical (anti)instanton solutions explicitly,
next we turn to the computation of their instanton number. Since we
have constructed the single instanton and anti-instanton solutions
above we anticipate that the assocated instanton number is $+1$ or $-1$
respectively. Recall that in the background of flat Euclidean spacetime,
the instanton number is equal to the ``Pontryagin index'' or the ``2nd
Chern class'' [1] given by
\begin{eqnarray}
\nu [A] = \int_{R^4}d^4x {-1\over 16\pi^2}Tr[F_{\mu\nu}\tilde{F}_{\mu\nu}]
\end{eqnarray}
where $Tr$ indicates the sum over repeated hidden group indices.
Therefore, for the case of pure YM gauge theory in the background of
curved spacetime like the present de Sitter case, one can similarly define 
the instanton number as the curved spacetime version of the Pontryagin
index. Thus we shall compute this curved spacetime version of the Pontryagin
index for our case. To do so, first note ;
\begin{eqnarray}
F^{a}_{\mu\nu}\tilde{F}^{a}_{\mu\nu} &=& F^{a}_{AB}\tilde{F}^{a}_{AB} =
2F^{a}_{0b}\tilde{F}^{a}_{0b} + F^{a}_{bc}\tilde{F}^{a}_{bc} \nonumber \\
&=& [2({H'\over Na})({H^2-1 \over a^2})\delta^{a}_{b}\delta^{b}_{a} +
({H^2-1 \over a^2})({H'\over Na})\epsilon^{abc}\epsilon_{abc}] \\
&=& {12\over Na^3}H'(H^2-1). \nonumber
\end{eqnarray}
Thus, the curved spacetime version of the Pontryagin index is
\begin{eqnarray}
\nu [A] &=& \int_{R\times S^3}d^4x\sqrt{g}{-1\over 32\pi^2}
[F^{a}_{\mu\nu}\tilde{F}^{a}_{\mu\nu}] \nonumber \\
&=& 2\pi^2 \int^{\infty}_{-\infty}d\tau Na^3 [{-1\over 32\pi^2}
\{{12\over Na^3}H'(H^2-1)\}] \\
&=& \pm {3\over 2}\int^{1}_{0}dH(H^2 - 1) \nonumber \\
&=& \pm 1 \nonumber
\end{eqnarray}
indicating that the instanton number is $+1$ for the single instanton 
solutions or $-1$ for the single anti-instanton solutions just as 
expected. \\
Now, as the final analysis of our classical instanton solution, we 
evaluate the instanton contribution to the vacuum-to-vacuum tunnelling
amplitude. The ``instanton action'', namely the Euclidean action
evaluated at the instanton is given by
\begin{eqnarray}
I_{YM}(instanton) &=& \int_{R\times S^3}d^4x \sqrt{g}[{1\over 4g^2_{c}}
(F^{a}_{\mu\nu})^2]\mid_{instanton} \nonumber \\
&=& {r^2_{0}\over 2}\int^{\infty}_{-\infty} d\tau [({a\over N})H'^2 +
({N\over a})(H^2 - 1)^2]\mid_{instanton} \nonumber \\
&=& r^2_{0}\int^{\infty}_{-\infty} d\tau ({N\over a})(H^2 - 1)^2 \\
&=& -2r^2_{0}\int^{1}_{0}dH (H^2 - 1) = {8\pi^2 \over g^2_{c}} \nonumber
\end{eqnarray}
where we used the (anti)self-dual equation satisfied by (anti)instanton
solution, $(dH/d\tau) = \mp (N/a)(H^2-1)$ and $r^2_{0} = {6\pi^2/g^2_{c}}$.
This instanton action is essentially the same as the soliton energy we
computed earlier in eq.(19). Consequently, the semiclassical approximation
to the vacuum-to-vacuum transition amplitude is given by
\begin{eqnarray}
(inter-vacua ~tunnelling ~amplitude) &\sim & \exp{[-I_{YM}(instanton)]} \\
&=& e^{-{8\pi^2 \over g^2_{c}}}. \nonumber
\end{eqnarray}
It is interesting to note that this instanton contribution to the inter-vacua
tunnelling amplitude for the pure YM theory formulated in the background of
de Sitter spacetime turns out to be the same as that for YM theory in the
usual flat spacetime [1].  
At this point, it seems appropriate to comment on the relationship between
our present work and the work of Charap and Duff [2]. Charap and Duff [2]
also considered Euclideanized Einstein-Yang-Mills theory (but in the
absence of the cosmological constant) and looked for classical solutions 
to the (anti)self-dual equation in the YM sector. Again, since the YM field
energy-momentum tensor vanishes for (anti)self-dual field strength in the
Euclidean signature, the YM field does not disturb the spacetime geometry.
Thus for the ``background'' geometry, they particularly took the
Schwarzschild spacetime which has SO(3)-isometry and looked for instanton
solutions which possess SO(3)-symmetry or O(4)-symmetry (having particularly
the decomposition, $O(4)\approx SU(2)\times SU(2)$). It turned out that
the resulting instanton solutions are characterized by the Pontryagin
index $\nu[A] = \pm 1$ and have the instanton action 
$I_{YM}(instanton) = 8\pi^2/ g^2_{c}$ which are the same as those for 
the instanton in flat space. Therefore the last point, namely that the 
curved spacetime version of instanton solutions carry the Pontryagin index
of $\pm 1$ and particularly that the inter-vacua tunnelling amplitude
remains the same even when the gravity of certain type is turned on are
precisely the same as the results of our study given in this section.

\centerline {\rm \bf IV. Fermionic zero modes in the instanton background}

In flat spacetime, one is usually interested in the dynamics of chiral
fermions in the instanton background in order to explore phenomenon like
chirality-changing fermion propagation due to the background instanton
configuration. Thus here we also consider the dynamics of chiral fermions
in the background of our curved (i.e., de Sitter) spacetime version of
instanton. Thus we begin by checking if there are fermionic zero modes
in this curved spacetime instanton background. \\
The Dirac equations for massless SU(2) doublet fermion field in this
curved spacetime instanton background are given by
\begin{eqnarray}
 &&\gamma^{C}e^{\mu}_{C}
 [\overrightarrow{\partial}_{\mu} 
 - {i\over4}\omega^{AB}_{\mu}\sigma_{AB} 
 - iA^{a}_{\mu}T^{a}]\Psi = 0, \\
 &&\bar{\Psi}\gamma^{C}e^{\mu}_{C}
 [\overleftarrow{\partial}_{\mu} 
 - {i\over4}\omega^{AB}_{\mu}\sigma_{AB} 
 - iA^{a}_{\mu}T^{a}] = 0 \nonumber
\end{eqnarray}
where the covariant derivative is given by
\begin{eqnarray}
 \gamma^{\mu}\nabla_{\mu}
= \gamma^{\mu}
     [\partial_{\mu} - {i\over4}\omega^{AB}_{\mu}\sigma_{AB}
        - iA^{a}_{\mu}T^{a}] \nonumber
\end{eqnarray}
where $e^{A}_{\mu}(x)\Bigl(e_{A}^{\mu}(x)\Bigr)$ is the ``vierbein'' 
(and it's inverse) defined by 
$g_{\mu\nu}(x) = \delta_{AB}e^{A}_{\mu}(x)e^{B}_{\nu}(x)$ 
and $e^{A}_{\mu}e_{B}^{\mu} = \delta^{A}_{\ B}$ , 
$e_{A}^{\mu}e^{A}_{\nu} = \delta^{\mu}_{\ \nu}$ 
and $e\equiv(det ~e^{A}_{\mu})$. Thus the Greek indices $\mu,\nu$ refer 
to coordinate basis while the Roman indices A,B = 0,1,2,3 refer 
to non-coordinate basis.  In addition, $\gamma^{\mu}(x) = e^{\mu}_{A}(x)\gamma^{A}$ 
is the curved spacetime $\gamma$-matrices obeying 
$\{\gamma^{\mu}(x),\gamma^{\nu}(x)\} = -2g_{\mu\nu}(x)$ 
with $\gamma^{A}$ being the usual flat spacetime $\gamma$-matrices.  
Next $(\partial_{\mu} - {i\over4}\omega_{\mu}^{AB}\sigma_{AB})$ is then 
the Lorentz covariant derivative with 
$\omega_{\mu B}^{A} = -e^{\nu}_{B}
 (\partial_{\mu}e_{\nu}^{A} - \Gamma^{\lambda}_{\mu\nu}e^{A}_{\lambda})$
being the spin connection and 
$\sigma^{AB} = {i\over2}[\gamma^{A},\gamma^{B}]$
being the $SO(3,1)$ group generator in the spinor representation. \\
Now, in order eventually to confirm the Atiyah-Patodi-Singer index
theorem [9] in our system under consideration, we begin with the brief
review of the ``mixed'' anomaly and the associated index theorem in
terms of Fujikawa's path integral formulation [10]. 
As usual, consider the chiral $U(1)_{A}$ transformation of the spinor
field coupled to both the background gauge and gravitational field
\begin{eqnarray}
\Psi(x) &\rightarrow& \Psi'(x) = e^{i\gamma_{5}\alpha(x)}\Psi(x), \\
\bar{\Psi}(x) &\rightarrow& \bar{\Psi}'(x) = \bar{\Psi}(x)
e^{i\gamma_{5}\alpha(x)}. \nonumber
\end{eqnarray}
Under this chiral transformation, the fermionic integration measure
in the functional integral
\begin{eqnarray}
Z = \int [d\Psi d\bar{\Psi}]e^{-S_{F}[\Psi, \bar{\Psi}]}
\end{eqnarray}
changes by the Jacobian determinant
\begin{eqnarray}
[d\Psi d\bar{\Psi}] \rightarrow [d\Psi' d\bar{\Psi}'] = J(\alpha)
[d\Psi d\bar{\Psi}]
\end{eqnarray}
with $J(\alpha) = \exp{[-i\int d^{2n}x \alpha(x)A(x)]}$ and as is
well-known, this leads to the mixed (gauge $+$ gravity) anomaly
\begin{eqnarray}
\nabla_{\mu}<\bar{\Psi}\gamma^{\mu}\gamma_{5}\Psi > = 
i<{1\over e}A(x)>
\end{eqnarray}
where $<...>$ stands for expectation value in terms of the functional
integral. And this ``anomaly term'' on the right hand side of eq.(40)
is related to the Dirac index
in the index theorem. Namely, the Atiyah-Patodi-Singer index theorem
states that the analytical index defined by
$({\rm index}\nabla)_{2n} = {\rm dim} {\rm ker}(\nabla) - 
{\rm dim} {\rm ker}(\nabla^{\dagger})$ is just a topological invariant
expressed in terms of an integral of an appropriate characteristic
class over the $2n$-dim. manifold $M^{2n}$, i.e.,
\begin{eqnarray}
({\rm index}\nabla)_{2n} &=& {\rm dim} {\rm ker}(\nabla) -
{\rm dim} {\rm ker}(\nabla^{\dagger}) \nonumber \\
&=& \int_{M^{2n}}\hat{A}(R)Ch(F) 
\end{eqnarray}
where $\hat{A}(R)$ and $Ch(F)$ denote ``$A$-roof (or Dirac) genus'' and
the ``total Chern character'' respectively defined by
\begin{eqnarray}
\hat{A}(R) &\equiv& \prod^{n}_{i=1} [{(x_{i}/2)\over \sinh (x_{i}/2)}],
\qquad (x_{i} \equiv {iR\over 2\pi}), \nonumber \\
Ch(F) &\equiv& Tr \exp{({iF\over 2\pi})} 
\end{eqnarray}
with $R$ and $F$ being curvature and YM field strength 2-forms
respectively. Generally in $2n$-dim., $\hat{A}(R)$ and $Ch(F)$ can be
expanded in series of Pontryagin classes and Chern classes respectively
and particularly in 4-dim., which is of our interest,
\begin{eqnarray}
({\rm index} \nabla)_{4} &=& {-1\over 8\pi^2}\int_{M^4}Tr(F\wedge F)
+ {1\over 192 \pi^2}\int_{M^4}Tr(R\wedge R) \nonumber \\
&=& {-1\over 16\pi^2}\int_{M^4}d^4x \sqrt{g}Tr(F_{\mu\nu}
\tilde{F}_{\mu\nu}) - {1\over 8}\tau(M^4) 
\end{eqnarray}
where $\tau(M) \equiv \int_{M}({-1\over 24 \pi^2})Tr(R\wedge R)$ denotes
the ``Hirzebruch signature'' of the manifold $M$. \\
For the case at hand, however, the background spacetime manifold $M^4$ is
the Euclidean de Sitter space with the geometry and topology of that of 
$S^4$ whose Hirzebruch signature is $\tau(S^4) = 0$. 
Therefore, for massless fermions
in the background of YM gauge field (particularly instantons) and in the
background of Euclidean de Sitter space, the appropriate form of the
Atiyah-Patodi-Singer index theorem reads
\begin{eqnarray}
({\rm index}\nabla)_{4}  
 = \int_{S^4}d^4x \sqrt{g}{-1\over 16\pi^2}Tr(F_{\mu\nu}
\tilde{F}_{\mu\nu}) = \pm 1
\end{eqnarray}
as has been evaluated earlier in eq.(33). Thus all we need to do is to
confirm this relation by checking if there actually is at least one
normalizable positive-chirality fermion zero mode or negative-chirality
fermion zero mode. And to see this, the most straightforward way is to
solve the Dirac equation for massless fermion field given earlier explicitly.
In order to solve the Dirac equation, we need explicit expressions for 
the spin connection $\omega^{AB}_{\mu}$ of de Sitter background
spacetime (represented by $k=+1$ FRW-metric) and the YM gauge connection
$A^{a}_{\mu}$ of the background instanton. \\
First, we can obtain the spin connection 1-forms, using the non-coordinate
basis 1-forms given in eq.(4) and the Cartan's 1st structure equation
(i.e., the torsion-free condition)
\begin{eqnarray}
de^{A} + \omega^{A}_{B}\wedge e^{B} = 0
\end{eqnarray}
along with the help of Maurer-Cartan structure equatin given in eq.(5).
And they are
\begin{eqnarray}
 \omega^{a}_{\mu0} 
 = -\omega^{0}_{\mu{a}} 
 = {1\over N}({a'\over{a}})e^{a}_{\mu}\quad,\quad
 \omega^{a}_{\mu{b}} 
 = -\omega^{b}_{\mu{a}} 
 = {-1\over a}\epsilon^{abc}e^{c}_{\mu}.
\end{eqnarray}
Next, the YM gauge connection 1-form can be given in this 
non-coordinate basis as well.  Namely, using
$A^{a} = A^{a}_{\mu}dx^{\mu} = A^{a}_{B}e^{B} = A^{a}_{b}e^{b}
= [(1+H)/a]e^{a}$ (since we chose the temporal gauge,
$A_{0} = 0$), we get
\begin{eqnarray}
A^{a}_{b} = [{1 + H(\tau) \over a(\tau)}]\delta^{a}_{b}.
\end{eqnarray}
Now we are ready to solve the Dirac equation in eq.(37) using the concrete
forms for the spin connection given in eq.(46) and for the YM gauge connection
given in eq.(47).
In addition, assuming that the fermion field depends only on the Euclidean time 
$\tau$ and setting 
\begin{eqnarray}
 \Psi(\tau) = a^{-{3\over2}}(\tau)\tilde{\Psi}(\tau),
\end{eqnarray}
the Dirac equation in eq.(37) reduces to
\begin{eqnarray}
[\partial_{\tau} - {N\over 4a}\epsilon_{abc}\gamma^{0}\gamma^{a}\gamma^{b}
\gamma^{c} - i\gamma^{0}\gamma^{a}T^{a}{N\over a}(1+H)]\tilde{\Psi}(\tau) = 0.
\end{eqnarray}
Further using $\epsilon_{abc}\gamma^{0}\gamma^{a}\gamma^{b}\gamma^{c} =
3!\gamma_{5}$ (where  $\gamma_{5} = \gamma^{0}\gamma^{1}\gamma^{2}\gamma^{3}$
is the Euclidean $\gamma_{5}$-matrix) and the original Dirac matrices
$\gamma^{0}=\beta$, $\gamma^{a}=\beta \alpha^{a}$ with matrices $\beta$,
$\alpha^{a}$ satisfying $\{\alpha^{a},~\alpha^{b}\} = 2\delta^{ab}$,
$\{\alpha^{a},~\beta\} = 0$ and $\beta^2 = I$, this Dirac equation can be
rewritten as
\begin{eqnarray}
[\partial_{\tau} - \gamma_{5}{3N\over 2a}
- i{1\over 2}\alpha^{a}\tau^{a}{N\over a}(1+H)]\tilde{\Psi}(\tau) = 0
\end{eqnarray}
whose solution is readily given as
\begin{eqnarray}
 \tilde{\Psi}(\tau) 
 = \exp{[{1\over 2}\int^{\tau} d\tau' {N\over a}  
 \{3\chi + i(\alpha^{a}\tau^{a})(1 + H)\}]} ~U
\end{eqnarray}
where $\chi =\pm 1$ depending on the constant basis spinor $U$ 
which may have  positive
chirality, $\gamma_{5}U_{+}=U_{+}$ or negative chirality, 
$\gamma_{5}U_{-}=-U_{-}$ respectively. Note that $[\gamma_{5},~\alpha^{a}]
= 0$ since $\{\gamma_{5},~\beta\} = 0$, $\{\gamma_{5},~\beta \alpha^{a}\} = 0$.
This implies that the chirality state of $U$-spinor exactly mirrors that of
$\tilde{\Psi}(\tau)$ or $\Psi (\tau)$.
Finally, the solution to the massless Dirac equation is found to be
\begin{eqnarray}
 \Psi(\tau) = {1\over{a^{3\over2}(\tau)}}\tilde{\Psi}(\tau).\nonumber
\end{eqnarray}
And of course in this expression for the solution to the Dirac equation,
the information of the background de Sitter space and the background
instanton configuration is given by
\begin{eqnarray}
a(\tau) = {1\over \kappa}\cos (\kappa \tau), \qquad H(\tau) = 
\mp \sin \kappa \tau \nonumber
\end{eqnarray}
for the gauge choice $N(\tau) = 1$ and
\begin{eqnarray}
a(\tau) = {1\over \kappa \cosh \tau}, \qquad H(\tau) =
\mp \tanh \tau \nonumber
\end{eqnarray}
for the gauge choice $N(\tau) = a(\tau)$ with the minus (plus) sign refering to
instanton (anti-instanton) respectively. Since both $a(\tau)$ and $H(\tau)$ are
bounded and oscillating functions of $\tau$ and the $\tau$-integration in the
exponent is finite due to the finite integration range, clearly these massless
solutions to the Dirac equation above are ``normalizable'' zero modes.
Finally, since the Atiyah-Patodi-Singer index theorem states that
\begin{eqnarray}
({\rm index}\nabla)_{4} &=& {\rm dim} {\rm ker}(\nabla^{\dagger}\nabla) -
{\rm dim} {\rm ker}(\nabla \nabla^{\dagger}) \nonumber \\
&=& n_{+} - n_{-} = \pm 1 
\end{eqnarray}
where we used eq.(44). Namely, the difference in the number of positive-chirality 
fermion zero 
modes ($n_{+}$) and that of negative-chirality fermion zero modes ($n_{-}$) cannot
be arbitrary but is fixed by the Pontryagin index representing the instanton
number.  Thus for our case, for instanton with $\nu[A] = 1$, there should be,
say, one positive-chirality fermion zero mode ($n_{+} = 1$) with no
negative-chirality zero mode ($n_{-} = 0$) while for anti-instanton with
$\nu[A] = -1$, there should be one negative-chirality fermion zero mode
 ($n_{-} = 1$) with no positive-chirality zero mode ($n_{+} = 0$). 
And in the above we have seen that this rule can indeed be obeyed since there
is only one normalizable zero mode solution which could have either positive
or negative chirality state. 

\centerline {\rm \bf V. Quantisation of instanton} 
{\bf (1) General description of soliton quantisation scheme} 
 
Before we carry out the explicit quantisation of the instanton in the YM theory
formulated in the background of de Sitter spacetime represented by $k = +1$
FRW-metric, we provide a brief review of the conventional soliton quantisation
scheme. Historically, the formalism for performing soliton quantisation has
been developed in the original papers through a variety of techniques [5,6] and
here in this work, we shall mainly refer to the formalism of Dashen et al. [5]
which is generally known to be standard. In general, solitons can be associated 
with quantum extended-particle states. And certain properties of these quantum
states like their energy, for instance, can be expanded in a semiclassical 
series. The leading terms in this series will be seen to be related to the
corresponding classical soliton solutions. In this fashion, knowledge of the
classical soliton solutions will yield some information about the quantum
particle states, in a systematic semiclassical expansion. Moreover, this
information will be non-perturbative in the non-linear couplings since, in
most cases, the corresponding classical solutions are themselves non-pertubative.
Now, in order to demonstrate, in a general manner, the quantisation of static 
soliton to obtain extended, non-perturbative, quantum particle states, we
consider a scalar field theory governed by the Lagrangian
\begin{eqnarray}
L = \int d^3x [{1\over 2}({\partial \phi \over \partial t})^2 - 
{1\over 2}(\nabla \phi)^2 - U(\phi)]
\end{eqnarray}
The classical 
dynamics of this system is quite similar to particle mechanics. For instance,
the Lagrangian has the familiar standard form $L = T[\phi] - V[\phi]$ with
the kinetic and potential energy being given by
\begin{eqnarray}
T[\phi] = \int d^3x {1\over 2}({\partial \phi \over \partial t})^2, 
~~~V[\phi] = \int d^3x [{1\over 2}(\nabla \phi)^2 + U(\phi)]. 
\end{eqnarray}
Next, extremizing this Lagrangian yields the following Euler-Lagrange's
equation of motion
\begin{eqnarray}
{\partial^2 \phi (t,\vec{x})\over \partial t^2} = - {\delta V[\phi]\over
\delta \phi (t,\vec{x})}
\end{eqnarray}
As a particle mechanical 
analogue, this classical field equation is similar to Newton's equation of
motion with the field $\phi(t,\vec{x})$ playing the role of the ``coordinates''.
First note that static solutions $\phi(t,\vec{x}) = \phi(\vec{x})$ satisfying
\begin{eqnarray}
{\delta V[\phi] \over \delta \phi (\vec{x})} = - 
{\delta L\over \delta \phi(\vec{x})} = 0
\end{eqnarray}
are automatically the extremum points for both the Lagrangian and the potential
energy $V[\phi ]$ in field space. In particular, stable static solutions such
as the vacuum or soliton solutions are ``minima'' of $V[\phi ]$ just as in
particle mechanics. Let $\phi(\vec{x}) = \phi_{0}(\vec{x})$ be one such 
minimum, then we can make a functional Taylor expansion of $V[\phi ]$ about
$\phi_{0}$ at which $V[\phi ]$ gets minimized ;
\begin{eqnarray}
V[\phi] = V[\phi_{0}] + \int d^3x {1\over2!}\{\eta(\vec{x})[-\nabla^2 +
({d^2U\over d\phi^2})\mid_{\phi_{0}(\vec{x})}]\eta(\vec{x}) + ...\}
\end{eqnarray}
where $\eta(\vec{x}) \equiv \phi(\vec{x}) - \phi_{0}(\vec{x})$, and integration
by parts has been used and `dots' represent cubic and higher terms. These
higher order terms would be small and thus can be neglected or treated in
perturbation provided the magnitude of the fluctuations $\eta(\vec{x})$ is
small and/or the third and higher derivatives of $V[\phi]$ at $\phi_{0}$
are small. Thus to lowest order in this approximation expansion,
eigenvalues and eigenfunctions of the operator
$[-\nabla^2 + (d^2U/ d\phi^2)\mid_{\phi_{0}}]$
will be given by the following differential equation
\begin{eqnarray}
[-\nabla^2 + ({d^2U\over d\phi^2})\mid_{\phi_{0}(\vec{x})}]\eta_{i}(\vec{x})
 = \omega^2_{i} \eta_{i}(\vec{x})
\end{eqnarray}
where $\eta_{i}(\vec{x})$ are the orthonormal ``normal modes'' of
fluctuations around $\phi_{0}(\vec{x})$. Then next, following Creutz [11],
introduce
\begin{eqnarray}
\eta(t,\vec{x}) \equiv  \phi(t,\vec{x}) - \phi_{0}(\vec{x})
\equiv  \sum_{i} C_{i}(t)\eta_{i}(\vec{x}). 
\end{eqnarray}
Then using the orthogonality, $\int d^3x \eta_{i}(\vec{x})\eta_{j}(\vec{x}) =
\delta_{ij}$, the Lagrangian of this system becomes
\begin{eqnarray}
L = {1\over 2}\sum_{i}[\dot{C}_{i}(t)]^2 - \left( V[\phi_{0}] + {1\over 2}
\sum_{i}[C_{i}(t)]^2 \omega^2_{i} \right) + ...
\end{eqnarray}
where $\dot{C}_{i} \equiv dC_{i}/dt $ and dots stand for contributions from 
higher terms. Evidently, this reduced Lagrangian represents that of a set of
harmonic oscillators, one for each normal mode, apart from a constant term
$V[\phi_{0}]$. 
For the sake of definiteness, if we take the usual $\phi^4$-theory as an
example, $U(\phi) = {1\over 2}m^2\phi^2 + {\lambda \over 4}\phi^4$. Then
applying the above general formulation with the choice of the stable static
solution $\phi_{0}(t,\vec{x}) = 0$ leads to $\eta{i}(\vec{x}) = 
{1\over \sqrt{L^3}}\exp{(i\vec{k}_{i}\cdot \vec{x})}$ and 
$\omega^2_{i} = (\vec{k}^2_{i} + m^2)$ with $k_{i}L = 2\pi N_{i}$ 
($L \rightarrow \infty$) in the box-normalization which amounts to a
quantisation condition. Correspondingly, in quantum theory one can construct 
a set of approximate harmonic oscillator states with energies given by
\begin{eqnarray}
E_{\{n_{i}\}} = V[\phi_{0}] + \hbar \sum_{i}(n_{i}+{1\over 2})
[\vec{k}^2_{i} + m^2]^{1/2}
\end{eqnarray}
where $n_{i}$ is the excitation number of the $i$-th normal mode.
It relates, approximately, the energies of certain quantum levels to the
classical solution $\phi_{0}(\vec{x})$. The second term involves $\omega_{i}$,
which are the stability frequencies of $\phi_{0}(\vec{x})$. This completes
the short review of the standard soliton quantisation scheme that we shall
employ in the case of our interest. 

{\bf (2) Quantisation of the instanton} 

Now, returning to our problem, consider the Euclidean action of our system,
i.e., the pure YM theory in de Sitter background spacetime given earlier,
\begin{eqnarray}
I_{YM} \equiv V[H] = {r^2_{0}\over 2}\int d\tilde{\tau} 
[({dH\over d\tilde{\tau}})^2 + (H^2 - 1)^2].
\end{eqnarray}
Since the ``Euclidean time'', $\tilde{\tau}$ is just another ``spacelike''
coordinate, this Euclidean action may be viewed as the potential energy
$V[H]$ or the Hamiltonian $H_{YM}$, in the above general formalism, of a
system of a scalar field $H(\tilde{\tau})$ with the ``double-well'' potential
$U(H) = {1\over 2}(H^2-1)^2$. Namely, our system can be viewed as a kind of
scalar $H^4$-theory in a stable static soliton sector. Therefore one can apply
the standard soliton quantization scheme described above to the quantisation
of vacuum and instantons of our theory. To be more specific, we would like
to explore the energy spectrums of excitations around the vacuum and the
instanton configurations. 

A. The vacuum and its excitations 

We first begin with the excitations around the classical vacuum (i.e.,
one of the two degenerate vacua), $H_{1}(\tau) = 1$. The Euclidean action
actually represents the energy of this system and it can be expanded around the 
vacuum $H_{1} = 1$ as 
\begin{eqnarray}
I_{YM}[H] = I_{YM}[H_{1}] &+& r^2_{0}\{\int d\tilde{\tau}{1\over 2}\tilde{H}
[-{\partial^2\over \partial \tilde{\tau}^2} + 4]\tilde{H} \\
&+& \int d\tilde{\tau} 2\tilde{H}^3 + \int d\tilde{\tau}{1\over 2}\tilde{H}^4
+ O(\tilde{H}^5)\} \nonumber
\end{eqnarray}
where $\tilde{H} \equiv (H - H_{1}) = H - 1$ and $I_{YM}[H_{1}]=0$. Note that
since $r^2_{0}$ is an overall factor commonly multiplied to all terms in the
Euclidean action (or the Hamiltonian), we will henceforth work with the 
rescaled Euclidean action $I'_{YM} = I_{YM}/r^2_{0}$ and then restore this
overall factor $r^2_{0}$ at the end of the computation of the energies.
Now, if we restrict our interest to excitations arising from sufficiently
small deviations from the classical vacuum, i.e., $\tilde{H}=(H-1)<<1$,
then terms higher than the cubic term can be ignored. Then in the lowest-
order quadratic term, the second functional derivative of $I'_{YM}[H]$
at $H = 1$ is the operator $[-{\partial^2 \over \partial \tilde{\tau}^2}
+ 4]$, whose eigenvalues are $(k^2_{n} + 4)$ with eigenfunctions
$e^{ik_{n}\tilde{\tau}}$. Then the allowed values of $k_{n}$ are
obtained, in box-normalization, by
\begin{eqnarray}
k_{n}L = 2\pi n
\end{eqnarray}
where $n\in Z$ and $L$, the length of the box, will ultimately tend to
infinity with the replacement,
\begin{eqnarray}
\sum_{k_{n}} \Longrightarrow L\int {dk\over (2\pi)}.
\end{eqnarray}
Now, we can construct a tower of approximate harmonic oscillator states
around the vacuum $H_{1} = 1$, the lowest of which has the energy,
restoring the overall factor $r^2_{0}$,
\begin{eqnarray}
E_{vac} \cong 0 + {1\over 2}\hbar r^2_{0}\sum_{n}[k^2_{n} + 2^2]^{1/2}
\end{eqnarray}
where the zero represents the classical vacuum energy $I_{YM}[H_{1}=1]$.
This is the quantum state of the vacuum of the system. Next, higher
excitations will have energies
\begin{eqnarray}
E_{vac} \cong \hbar r^2_{0} \sum_{n}(N_{n}+{1\over 2})[k^2_{n} + 2^2]^{1/2}.
\end{eqnarray}
These correspond to the familiar quanta of the theory, where $N_{n}$ of
them have momentum $\hbar k_{n}$. We will call this set of states built
around the vacuum $H_{1} = 1$, the ``vacuum sector''. Since this procedure
essentially quantizes the shifted field, $\tilde{H} = (H-H_{1}) = (H-1)$, 
as in standard perturbation methods, we can borrow the familiar result 
to lowest order that
\begin{eqnarray}
<0|\tilde{H}(\tau)|0> = 0 \qquad {\rm or} \qquad <0|H(\tau)|0> = H_{1} = 1
\end{eqnarray}
where $|0>$ denotes the vacum state. 

B. The quantum instanton and its excitations 

Next, we turn to the excitations around the instanton configuration. For 
the sake of definiteness, we choose to work with the instanton solution
resulting from the gauge fixing $N(\tau) = a(\tau)$
\begin{eqnarray}
H_{c}(\tau) = \tanh \tau \nonumber
\end{eqnarray}
with energy (i.e., Euclidean action evaluated at this instanton solution)
\begin{eqnarray}
I_{YM}[H_{c}] = {8\pi^2 \over g^2_{c}}. \nonumber
\end{eqnarray}
Clearly, this instanton solution is an extremum point of the Euclidean 
action $I_{YM}[H]$. Thus again, the Euclidean action, namely the energy
of the system can be expanded around this instanton solution, i.e., the
extremum point $H_{c}(\tau)$ as
\begin{eqnarray}
I_{YM}[H] = I_{YM}[H_{c}] &+& r^2_{0}\{\int d\tilde{\tau}{1\over 2}\tilde{H}
[-{\partial^2\over \partial \tilde{\tau}^2} - 2 + 6H^2_{c}]\tilde{H} \\
&+& \int d\tilde{\tau} [2H_{c}\tilde{H}^3 + {1\over 2}\tilde{H}^4]
+ O(\tilde{H}^5)\} \nonumber
\end{eqnarray}
where now $\tilde{H} \equiv (H - H_{c})$ and $d\tilde{\tau} = d\tau (N/a) =
d\tau$ since this instanton solution corresponds to the gauge choice
$N(\tau) = a(\tau)$. Henceforth, again, we shall work with the rescaled
Euclidean action $I'_{YM} = I_{YM}/r^2_{0}$. In the lowest-order
quadratic term, the eigenvalues of the second functional derivative of 
$I_{YM}[H]$ at $H_{c}$ are given by the equation
\begin{eqnarray}
[&-&{\partial^2\over \partial \tau^2} - 2 + 6H^2_{c}]\tilde{H}_{n}
(\tau) \\
&=& [-{\partial^2\over \partial \tau^2} - 2 + 6\tanh^2\tau]\tilde{H}_{n}
(\tau) = \omega^2_{n}\tilde{H}_{n}(\tau). \nonumber
\end{eqnarray}
Dividing this equation through by $2$, it becomes a Schr\H odinger-type
equation
\begin{eqnarray}
[-{1\over 2}{\partial^2\over \partial \tau^2} + (3\tanh^2 \tau - 1)]
\tilde{H}_{n}(\tau) = {\omega^2_{n}\over 2}\tilde{H}_{n}(\tau).
\end{eqnarray}
Fortunately, the eigenfunctions and eigenvalues of this Schr\H odinger-type
equation are exactly known [12]. It has two discrete levels followed by a
continuum. The discrete levels are ;
\begin{eqnarray}
\omega^2_{0} &=& 0 \qquad {\rm with} \qquad \tilde{H}_{0}(\tau) =
{1\over \cosh^2 \tau} \\
\omega^2_{1} &=& 3 \qquad {\rm with} \qquad \tilde{H}_{1}(\tau) =
{\sinh \over \cosh^2 \tau} \nonumber
\end{eqnarray}
This is followed by a continuum of levels which we shall label by $q$
rather than by $n\geq 2$. These are ;
\begin{eqnarray}
\omega^2_{q} = [q^2 + 2^2] \qquad {\rm with} \qquad
\tilde{H}_{q}(\tau) = e^{iq\tau}[3\tanh^2\tau - 1 - q^2 - 3iq\tanh\tau].
\end{eqnarray}
Here, the allowed values of $q$, like the allowed values of $k_{n}$ in the
vacuum case, are fixed by periodic boundary conditions in a box of length
$L$, with $L\rightarrow \infty$. It is noteworthy that the quantum 
fluctuation or excitation around the classical instanton solution
$\tilde{H}_{q}(\tau)$ above has an asymptotic behavior
\begin{eqnarray}
\tilde{H}_{q}(\tau) \longrightarrow \exp{[i(q\tau \pm {1\over 2}\delta(q))]}
\qquad {\rm as} ~~\tau \rightarrow \pm \infty \nonumber 
\end{eqnarray}
where $\delta(q) = - 2\arctan [3q/(2-q^2)]$ is just the phase shift of the
scattering states of the associated Schr\H odinger problem above. This is
precisely the quantum fluctuation around the classical vacuum,
\begin{eqnarray}
\tilde{H}_{n}(\tau) = e^{ik_{n}\tau} \qquad {\rm with} \qquad
\omega^2_{n} = [k^2_{n} + 2^2] \nonumber
\end{eqnarray}
we obtained earlier modulo phase shift just as expected since $\tau \rightarrow
\pm \infty$ is the vacuum limit. Now, as before, the allowed values of $q$ are
determined by thr periodic boundary condition in box-normalization
\begin{eqnarray}
q_{n}L + \delta(q_{n}) = 2\pi n, \qquad n\in Z.
\end{eqnarray}
In the $L\rightarrow \infty$ limit, these allowed values merge into a 
continuum with the replacement
\begin{eqnarray}
\sum_{q_{n}} \Longrightarrow \int^{\infty}_{-\infty}{dq\over (2\pi )}
[L + {\partial\over \partial q}\delta(q)].
\end{eqnarray}
Now we are ready to write down (or construct) the energy spectrum of
quantized instanton as a sum of the energy of the classical instanton and
the energy levels of the small quantum fluctuations (excitations) around
that classical instanton. Notice that in the expansion of the Euclidean
action, i.e., the energy of the system around the instanton, namely its
extremum point $H_{c}(\tau)$, if we restrict our interest to excitations
arising from sufficiently small deviation from the classical instanton,
i.e., $\tilde{H} \equiv (H - H_{c}) << 1$, terms higher than the cubic term
can be neglected and we are left with the minimum energy (i.e., energy
of the classical instanton) and quadratic term in $\tilde{H}(\tau)$.
Then this lowest-order quadratic term can be identified with the one
representing energy levels of a set of approximate harmonic oscillator
states spread in field space around $H_{c}(\tau)$ with the neglected 
higher-order terms representing all anharmonic terms. Therefore, in
this approximation of quantum fluctuations around the classical instanton
by a set of harmonic oscillator states in field space, the energy
spectrum of quantized instanton is given by (restoring the overall 
factor $r^2_{0}$)
\begin{eqnarray}
E_{\{N_{n}\}} &\cong & I_{YM}[H_{c}] + \hbar r^2_{0}\sum^{\infty}_{n=0}
(N_{n} + {1\over 2})\omega_{n} \\
&=& {4\over 3}r^2_{0} + (N_{1}+{1\over 2})\hbar r^2_{0}\sqrt{3} +
\hbar r^2_{0}\sum_{q_{n}} (N_{q_{n}}+{1\over 2})[q^2_{n}+2^2]^{1/2}
\nonumber
\end{eqnarray}
where, as we did in the quantized vacuum case, we explicitly retain
$\hbar$ for a few steps since we wish to bring out the semiclassical
nature of our theory. There is, however, a point to which one should 
be cautious ; while this analysis used in the treatment of small
quantum fluctuations around the classical instanton is essentially
valid for all the $n\geq 1$ modes, it does not hold for the $n = 0$
mode, because $\omega_{0} = 0$. Namely, unlike the $n\ge 1$ modes
which are genuine vibrational modes for small fluctuations, the
$n= 0$ mode is not vibrational at all. The `spring constant'
$\omega_{0}$ vanishes. Correspondingly, the quantum wave function
along the $n=0$ mode will not be confined near a given classical
solution, but will tend to spread. Now we turn to the interpretation
of the energy spectrum of the tower of quantized instanton states
given above ; 

(i) The lowest-energy state, 
\begin{eqnarray}
E_{0} \equiv E_{\{N_{n}=0\}} = {4\over 3}r^2_{0} + {1\over 2}\hbar r^2_{0}
\sqrt{3} + {1\over 2}\hbar r^2_{0}\sum_{q_{n}}[q^2_{n}+2^2]^{1/2}
\end{eqnarray}
may be interpreted as the ``lowest energy state of the quantum instanton''.
Note that although this state has lowest energy in the `instanton sector',
obviously it is not the absolute ground state or vacuum of this theory.
As we already discussed earlier, the vacuum of this theory has been
identified as the lowest energy state in the `vacuum sector',
\begin{eqnarray}
E_{vac} \cong {1\over 2}\hbar r^2_{0}\sum_{n}[k^2_{n} + 2^2]^{1/2}.
\end{eqnarray}
(ii) The next higher energy level, 
\begin{eqnarray}
E_{1} \equiv E_{\{N_{1}=1;N_{q_{n}}=0\}} = E_{0} + \hbar r^2_{0}\sqrt{3}
\end{eqnarray}
may be interpreted as a discrete excited state of the quantum instanton.
And higher excitations of this mode (i.e., $N_{1}>1$) give higher excited
states of the quantum instanton. 

(iii) The remaining higher energy states obtained by exciting the $n\geq 2$
modes (i.e., the $N_{q}\neq 0$ states) can be thought of as scattering
states ``quanta in the vacuum sector'' in the presence of the quantum
instanton. 

This is our interpretation of the family of quantum instanton states 
constructed around the classical instanton solution.  
Next, since we have constructed the quantum instanton states (i.e.,
small quantum fluctuations around the classical instanton solution)
and the associated energy spectrum, naturally the next question we
would like to ask is ; what would the effects of this quantum instanton
on the vacuum-to-vacuum tunnelling amplitude be ? Namely, we would like
to explore the lowest quantum correction to the Euclidean action and
hence to the vacuum-to-vacuum tunnelling amplitude arising from the
quantisation of the instanton. Since the saddle point (instanton)
approximation to the inter-vacua tunnelling amplitude is given by
\begin{eqnarray}
\Gamma \sim e^{-I_{YM}[instanton]},
\end{eqnarray}
one can naively expect that the tunnelling amplitude involving the
quantum correction coming from the contribution from the quantum
instanton at its lowest energy state (restoring the overall factor
$r^2_{0}$)
\begin{eqnarray}
E_{0} = {4\over 3}r^2_{0} + {1\over 2}\hbar r^2_{0}\sqrt{3} +
{1\over 2}\hbar r^2_{0}\sum_{q_{n}}[q^2_{n} + 2^2]^{1/2} \nonumber
\end{eqnarray}
would be given by 
\begin{eqnarray}
\Gamma \sim e^{-E_{0}}. \nonumber
\end{eqnarray}
Unfortunately, however, this naive prescription fails since the
expression for the lowest energy above is formally divergent. The infinite
series over $\sum_{q_{n}}$ in the last term of $E_{0}$ above
becomes, in the continuum limit,
\begin{eqnarray}
\int^{\infty}_{-\infty} {dq\over 2\pi}[L + {\partial \over \partial q}
\delta(q)] [q^2 + 2^2]^{1/2} \nonumber
\end{eqnarray}
i.e., a quadratically-divergent integral.
This in itself, however, need not worry us since the lowest energy of 
the quantum vacuum state is also quadratically divergent again when
the infinite series over $\sum_{k_{n}}$ is taken over to the continuum
limit, viz.,
\begin{eqnarray}
E_{vac} &\cong & {1\over 2}\hbar r^2_{0}\sum_{n}[k^2_{n} + 2^2]^{1/2} \\
&=& {1\over 2}\hbar r^2_{0}L \int^{\infty}_{-\infty}{dk\over (2\pi)}
[k^2 + 2^2]^{1/ 2} \rightarrow \infty. \nonumber
\end{eqnarray}
After all, what matters physically is the difference in energy between
any given state and the vacuum state. And this difference is obtained
by subtracting $E_{vac}$ from $E_{0}$ (From now on we will discuss the
regularization and the renormalization of the lowest energy of the
quantized instanton state. Since the overall factor $r^2_{0}$ is irrelevant
in the regularization procedure, we will henceforth work with the rescaled
energy $E' = E/r^2_{0}$ associated with the rescaled Euclidean action
$I'_{YM} = I_{YM}/r^2_{0}$. And at the end of the computations, we will
restore the overall factor $r^2_{0}$ in the final expression for the
renormalized value of the lowest energy of the quantized instanton.),
\begin{eqnarray}
E'_{0}-E'_{vac} = {4\over 3} + {1\over 2}\hbar \sqrt{3} + {1\over 2}\hbar
\sum_{n}[(q^2_{n}+4)^{1/2} - (k^2_{n}+4)^{1/2}].
\end{eqnarray}
Since both terms in the bracket are divergent, we must subtract them
carefully so as not to lose finite pieces. Let us start with a finite
box with size $L$. As usual, the periodic boundary condition in the
box-normalization determines the allowed values of $k_{n}$ and $q_{n}$
as
\begin{eqnarray}
2\pi n = k_{n}L = q_{n}L + \delta(q_{n}).
\end{eqnarray}
Thus, the term in bracket in ($E'_{0}-E'_{vac}$) becomes
\begin{eqnarray}
\{[(k_{n}-{\delta_{n}\over L})^{2}+4]^{1/2} - [k^2_{n}+4]^{1/2}\} =
- ({k_{n}\delta_{n}\over L})(k^2_{n}+4)^{-1/2} + O({1\over L^2})
\end{eqnarray}
where $\delta_{n} \equiv \delta(q_{n})$. 
Now going to the $L \rightarrow \infty$ limit and using the replacement
\begin{eqnarray}
\sum_{k_{n}}\Longrightarrow L\int {dk\over (2\pi)}, \nonumber
\end{eqnarray}
we have 
\begin{eqnarray}
E'_{0}-E'_{vac} = {4\over 3} + {1\over 2}\hbar \sqrt{3} - {\hbar \over 4\pi}
\int^{\infty}_{-\infty}dk {k\delta(k) \over \sqrt{k^2+4}}
\end{eqnarray}
where using $\delta(q) = -2\arctan [{3q/ (2-q^2)}]$ and 
$q_{n} = (k_{n} - \delta_{n}/L)$,
\begin{eqnarray}
\delta(k) = -2\arctan[{3k\over (2-k^2)}] + O({1\over L}).
\end{eqnarray}
Then next upon integrating by parts, we get
\begin{eqnarray}
E'_{0}-E'_{vac} &=& {4\over 3} + {1\over 2}\hbar \sqrt{3} - {\hbar \over 4\pi}
[\delta(k)\sqrt{k^2+4}]^{\infty}_{-\infty} 
+ {\hbar \over 4\pi}\int^{\infty}_{-\infty}dk\sqrt{k^2+4}{d\over dk}[\delta(k)]
\nonumber \\
&=& {4\over 3} + {1\over 2}\hbar \sqrt{3} - {3\hbar \pi} 
- {3\hbar \over 2\pi}\int^{\infty}_{-\infty}dk{(k^2+2)\over \sqrt{k^2+4}
(k^2+1)} 
\end{eqnarray}
where we used the phase shift in eq.(86). Now, although the quadratic divergence
in $E'_{0}$ has been removed by subtracting out $E'_{vac}$, $(E'_{0}-E'_{vac})$
still has a logarithmic divergence in the last term involving integral. In fact,
this divergence at this stage of the calculation need not concern us. We actually
should expect it to be there and it can be removed by ``normal ordering'' the
Hamiltonian. The occurrence of ultraviolet divergences in quantum field theory
due to the short-distance behavior of products of field operators is well-known
in standard perturbation theory. And typically, these divergences are removed 
by adding suitable ``counter terms'' to the Hamiltonian. Now for our theory,
the Euclidean action which is equivalent to the Hamiltonian is
\begin{eqnarray}
I'_{YM} = H'_{YM} = {1\over 2}\int d\tilde{\tau}[({dH\over d\tilde{\tau}})^2
+ H^4 - 2H^2 + 1].
\end{eqnarray}
In the quantised theory, operators like $H^2(\tau)$, $H^4(\tau)$ etc. are
formally divergent and ill-defined and thus so is the Hamiltonian. As a
consequence, energy levels calculated naively from this Hamiltonian will be
divergent as well. And this is the very reason behind the divergence in 
$(E'_{0}-E'_{vac})$. Now the removal of such divergences can be accomplished
by replacing the Hamiltonian by its normal ordered form $:H':$. In our
semiclassical formulation, however, it would be difficult to work directly
with the normal-ordered form. Instead, using the results of Wick's theorem, the 
normal-ordered form can be written as the original non-ordered form plus 
some counter terms,
\begin{eqnarray}
:H^4: &=& H^4 - AH^2 - B, \\
:H^2: &=& H^2 - C \nonumber
\end{eqnarray}
where $A$,$B$ and $C$ are constants which diverge in perturbation theory.
Therefore, the normal-ordered Hamiltonian may be written as (setting 
$A \equiv \partial m^2$ and $D \equiv B - 2C$)
\begin{eqnarray}
:H'_{YM}: &=& H'_{YM} - \int^{\infty}_{-\infty}d\tilde{\tau}[\partial m^2
H^2 + D] \nonumber \\
&\equiv & H'_{YM} + \Delta E' 
\end{eqnarray}
where the constants $\partial m^2$ and $D$ may be evaluated in perturbation
theory by standard methods. In particular, $\partial m^2$ is
the renormalization constant in the mass renormalization and to 1-loop
order, it is given by
\begin{eqnarray}
\partial m^2 = {12\hbar \over 16\pi}\int^{\Lambda}_{-\Lambda} {dk\over
\sqrt{k^2 - m^2}} = {3\hbar \over 4\pi}\int^{\Lambda}_{-\Lambda} {dk\over
\sqrt{k^2 + 2}}
\end{eqnarray}
where the numerical factor $12$ comes from the combinatorial factor of
associating each scalar field operator in $H^4(\tau)$ each with line in
the Feynman diagram and $\Lambda$ is the momentum cut-off. Also we used
the fact that in our scalar $H(\tau)$-field system represented by the 
Euclidean action or the Hamiltonian given in eq.(90), the mass squared
corresponds to $m^2 = -2$ and finally the factor $\hbar$ represents the 
1-loop correction. Next, we will not evaluate the other renormalization
constant $D$ since $(E'_{0}-E'_{vac})$ involves the difference between
two energy levels where the effect of $D$ will cancel out. 
Now we are ready to demonstrate that the counter term
\begin{eqnarray}
\Delta E' = -\int^{\infty}_{-\infty}d\tilde{\tau}[\partial m^2 H^2 + D]
\nonumber
\end{eqnarray}
indeed removes the logarithmic divergence in $(E'_{0}-E'_{vac})$.
Since the replacement of the Hamiltonian (or the Euclidean action)
$H'_{YM}$ by its normal ordered form $:H'_{YM}:$ amounts to adding
the counter term $\Delta E'$ above, in order to renormalize the lowest
energy of the quantum instanton state  $(E'_{0}-E'_{vac})$, we have
to add it by the counter terms
\begin{eqnarray}
(\Delta E'_{0}-\Delta E'_{vac}) &=& -\int^{\infty}_{-\infty}d\tilde{\tau}
[\partial m^2 H^2_{c}(\tilde{\tau})+D] + \int^{\infty}_{-\infty}d\tilde{\tau}
[\partial m^2 H^2_{1}(\tilde{\tau})+D] \nonumber \\
&=& (\partial m^2)\int^{\infty}_{-\infty}d\tau [1 - \tanh^2 \tau] \\
&=& 2(\partial m^2) = {3\hbar \over 2\pi}\int^{\Lambda}_{-\Lambda}
{dk\over \sqrt{k^2 + 2}}. \nonumber
\end{eqnarray}
Actually, this is the leading contribution of the counter terms to
 $(E'_{0}-E'_{vac})$ since we inserted classical instanton $H_{c}(\tau)$
and classical vacuum $H_{1}(\tau)$ into the ``quantum'' field $H(\tau)$
appearing in the counter term $\Delta E'$ above. Therefore finally, the finite,
renormalized lowest energy of the quantum instanton state is given by [5],
upon restoring the overall factor $r^2_{0}$,
\begin{eqnarray}
E^{ren}_{0} &\equiv & (E_{0}-E_{vac}) + (\Delta E_{0}-\Delta E_{vac}) 
\nonumber \\
&=& {4\over 3}r^2_{0} + \hbar r^2_{0}({\sqrt{3}\over 2} - {3\over \pi})
- {3\hbar \over 2\pi}r^2_{0}\int^{\infty}_{-\infty}dk
[{(k^2+2)\over \sqrt{k^2+4}(k^2+1)} - {1\over \sqrt{k^2+2}}] \nonumber \\
&=& {4\over 3}r^2_{0} + \hbar r^2_{0}({\sqrt{3}\over 6} - {3\over \pi})  
+ O(\hbar^3). 
\end{eqnarray}
Note here that both terms in the integrand behave as $1/k$ as 
$k\rightarrow \infty$ so that the logarithmic divergences cancel out.
To conclude, the ``renormalized'' (i.e., free of infinities of all sorts)
lowest energy of the quantized instanton state is given by
\begin{eqnarray}
E^{ren}_{0} = r^2_{0}[{4\over 3} - \hbar ({3\over \pi} - 
{\sqrt{3}\over 6})] 
= {8\pi^2 \over g^2_{c}}[1 - \hbar {3\over 4}({3\over \pi} -
{\sqrt{3}\over 6})].
\end{eqnarray}
We now make a few remarks on the renormalized value of the lowest energy
of the quantum instanton state given above. The first term $4r^2_{0}/3 =
8\pi^2/g^2_{c}$ is the Euclidean action evaluated at the classical
instanton solution i.e., the energy of the classical instanton. The next
term represents the leading correction coming from quantum fluctuations.
Thus appropriately, the first term is of order $\hbar^{0}$ and the second
term is of order $\hbar^{1}$. Secondly, in the weak-coupling limit,
$g_{c}<<1$, thanks mainly to the energy of the classical instanton, this
lowest energy of the quantised instanton is much larger than the lowest
energy of the quanta in the vacuum sector which is of order of 
$\hbar r^2_{0} = \hbar 6\pi^2/g^2_{c}$. 
In addition, it seems worth mentioning that what we have done so far
to get eq.(94) is the renormalization of the energy $E_{0}$ of the 
quantized instanton, {\it not} the usual renormalization of the 
gauge coupling constant $g_{c}$ of the theory. 
Finally we return to our major concern, namely the computation of the
lowest quantum correction to the Euclidean action and hence to the
vacuum-to-vacuum tunnelling amplitude arising from the quantisation of
the instanton. Note that the renormalized lowest energy of the quantized
instanton state $E^{ren}_{0}$ given above is {\it lower} than the energy
(Euclidean action) of the classical instanton, i.e.,
\begin{eqnarray}
E^{ren}_{0} = {8\pi^2 \over g^2_{c}}[1 - \hbar {3\over 4}({3\over \pi}
- {\sqrt{3}\over 6})] < I_{YM}[instanton] = {8\pi^2 \over g^2_{c}}
\end{eqnarray}
which then implies that
\begin{eqnarray}
\Gamma_{QI} \sim e^{-E^{ren}_{0}} > \Gamma_{CI} \sim e^{-I_{YM}[instanton]}
\end{eqnarray}
namely the inter-vacua tunnelling amplitude gets {\it enhanced} upon 
quantizing the instanton. Obviously, this is an expected result since the 
tunnelling between degenerate vacua is really a quantum phenomenon in
nature. Notice that our system, i.e., the pure YM theory in the background
of de Sitter spacetime represented by $k=+1$ FRW-metric exhibits, albeit
in a much simpler structure, almost all of the features of the YM theory
in flat spacetime as being unchanged including particularly the same
vacuum-to-vacuum tunnelling amplitude in the instanton approximation.
Thus we believe that the estimate of the lowest quantum correction to the
Euclidean instanton action and hence to the vacuum-to-vacuum tunnelling 
amplitude arising from the quantisation of the instanton would remain
the same even if we ask the same question in the pure YM theory in
flat spacetime although the actual computation of the corresponding
quantity (i.e., $E^{ren}_{0}$) would be even more formidable there!

\centerline {\rm \bf VI. Discussions}

We now summarize the results of the present work. In this work, we
examined, in detail, the instantons and their quantisation in pure
YM theory formulated in the background of de Sitter spacetime 
represented by spatially-closed ($k=+1$) FRW-metric. The SO(4)-symmetry
of the $k=+1$ FRW-metric having the topology of $S^4$ and
hence that of the dynamical YM field put on it, effectively reduced the
system to that of an one-dimensional self-interacting scalar field
theory with double-well potential. Since the vacuum structure of
this reduced system has just two-fold degeneracy and thus is relatively
simple, the associated instanton physics could be analyzed in a
quantitative manner. Classical instanton configurations have been
obtained as explicit solutions to the (anti)self-dual equation
which implies the Euclidean YM equation of motion. In addition,
the Pontryagin index representing the instanton number and the
inter-vacua tunnelling amplitude associated with these instanton
solutions have been evaluated. The single instanton and the single
anti-instanton are found to possess Pontryagin index $+1$ and $-1$
respectively, as expected. In particular, it is remarkable that
the semiclassical approximation (involving only the instanton
contribution) to the vacuum-to-vacuum tunnelling amplitude for
our YM theory formulated in de Sitter background spacetime turned
out to be the same as that for YM theory in the usual flat
spacetime. Atiyah-Patodi-Singer index theorem was also checked
in our system by demonstrating explicitly that there is only one
normalizable fermion zero mode in this de Sitter spacetime 
instanton background which has either positive or negative
chirality state. Lastly, we attempted the quantisation of our
instanton solution using the fact that the action or the
Hamiltonian of our reduced one-dimensional system takes on
precisely the same structure as that of one-dimensional scalar
field theory which admits kink soliton solutions.
Therefore, following the kink quantisation programme originally
proposed by Dashen, Hasslacher and Neveu, we performed the
quantisation of the vacuum and the instanton of our theory.
Of particular interest was the estimation of the lowest
quantum correction to the Euclidean action and hence to the
vacuum-to-vacuum tunnelling amplitude arising from the
quantisation of the instanton. It turned out that the
renormalized lowest energy of the quantized instanton state
is lower than the energy of the classical instanton. As a
consequence, the inter-vacua tunnelling amplitude gets
enhanced upon quantizing the instanton.

\centerline {\rm \bf Acknowledgements}

This work was supported in part by Korea Research Foundation and by Basic
Science Research Institute (BSRI-97-2427) at Ewha Women's Univ.

\newpage
\vspace*{2cm}

\centerline {\bf \large References}

\begin{description}

\item {[1]} See, for instance, R. Rajaraman, {\it Solitons and Instantons}
            (North-Holland, Elsevier Science Publishers, 1982).
\item {[2]} T. Eguchi and P. G. O. Freund, Phys. Rev. Lett. {\bf 37}, 1251
            (1977) ; J. M. Charap and M. J. Duff, Phys. Lett. {\bf B69},
            445 (1977) ; {\it ibid} {\bf B71}, 219 (1977). 
\item {[3]} P. Candelas and D. J. Raine, Phys. Rev. {\bf D12}, 965 (1975) ;
            J. S. Dowker and R. Critchley, Phys. Rev. {\bf D13}, 224 (1976) ;
            {\it ibid}, {\bf D13}, 3224 (1976) ;
            B. Ratra, Phys. Rev. {\bf D31}, 1931 (1985) ;
            B. Allen, Phys. Rev. {\bf D32}, 3136 (1985).              
\item {[4]} G. `tHooft, Nucl. Phys. {\bf B79}, 276 (1974) ;
            A. M. Polyakov, Pisma v. Zh. E.T.F., {\bf 20}, 430 (1974).
\item {[5]} R. F. Dashen, B. Hasslacher, and A. Neveu, Phys. Rev. {\bf D10},
            4130 (1974).            
\item {[6]} K. Cahill, Phys. Lett. {\bf B53}, 174 (1974) ;
            J. Goldstone and R. Jackiw, Phys. Rev. {\bf D11}, 1486 (1975) ;
            A. M. Polyakov, Phys. Lett. {\bf B59}, 82 (1975) ;
            N. H. Christ and T. D. Lee, Phys. Rev. {\bf D12}, 1606 (1975).
\item {[7]} A. Hosoya and W. Ogura, Phys. Lett. {\bf B22}, 117 (1989) ;
            S. -J. Rey, Nucl. Phys. {\bf B336}, 146 (1990).
\item {[8]} T. T. Wu and C. N. Yang, Phys. Rev. {\bf D12}, 3845 (1975).

\item {[9]} M. Atiyah and I. Singer, Ann. Math. {\bf 87}, 485 (1968) ;
            M. Atiyah, V. Patodi, and I. Singer, Math. Proc. Camb. Phil. Soc.,
            {\bf 79}, 71 (1976).

\item {[10]} K. Fujikawa, Phys. Rev. {\bf D21}, 2848 (1980).
            
\item {[11]} M. Creutz, Phys. Rev. {\bf D12}, 3126 (1975).

\item {[12]} P. Morse and H. Feshbach, {\it Methods of Mathmatical Physics}
             (McGraw-Hill Book Co., New York, 1953).

\end{description}

\end{document}